\documentclass[a4paper,fleqn,usenatbib,dvipdfmx]{mnras}
\usepackage{newtxtext,newtxmath}

\usepackage[T1]{fontenc}
\usepackage{ae,aecompl}

\usepackage{graphicx}	
\usepackage{amsmath}	
\usepackage{amssymb}	
\usepackage{bm}
\usepackage{subfigure}
\usepackage{color}

\title[Probing supervoids with weak lensing]{Probing supervoids with weak lensing}

\author[Y.Higuchi and K.T.Inoue]{
Yuichi Higuchi,$^{1}$\thanks{E-mail: yhiguchi@asiaa.sinica.edu.tw}
Kaiki Taro Inoue$^{2}$
\\
$^{1}$Academia Sinica Institute of Astronomy and Astrophysics (ASIAA), No. 1, Section 4, Roosevelt Rd, Taipei 10617, Taiwan\\
$^{2}$Faculty of Science and Engineering, Kindai University, Higashi-Osaka, Osaka, 577-8502, Japan\\
}

\date{Accepted XXX. Received YYY; in original form 2017}

\pubyear{2017}

\begin{document}
\label{firstpage}
\pagerange{\pageref{firstpage}--\pageref{lastpage}}
\maketitle

\begin{abstract}
The cosmic microwave background (CMB) has non-Gaussian features in the temperature fluctuations. An anomalous cold spot surrounded with a hot ring, called the Cold Spot is one of such features.
If a large underdence region (supervoid) resides towards the Cold Spot,
 we would be able to detect a systematic shape distortion in the images
 of background source galaxies via weak lensing effect.
In order to estimate the detectability of such signals, we used 
the data of $N$-body simulations to simulate full-sky ray-tracing of source galaxies. 
We searched for a most prominent underdense region using the simulated
 convergence maps smoothed at a scale of 20 degree and obtained
 tangential shears around it. The lensing signal expected in a concordant $\Lambda$CDM model can be detected at a signal-to-noise ratio $S/N\sim3$. If a supervoid with a radius of $\sim 200\,h^{-1}\,\textrm{Mpc}$ and a
 density contrast $\delta_0 \sim -0.3$ at the centre resides at a redshift $z\sim 0.2$,
 on-going and near-future weak gravitational lensing surveys would
 detect a lensing signal with $S/N\gtrsim4$ without resorting to stacking. 
From the tangential shear profile, we can obtain a constraint on the projected mass distribution of the supervoid.
\end{abstract}




\section{Introduction}
While most outcomes from the cosmic microwave background (CMB)
experiments such as {\it Wilkinson Microwave Anisotropy Probe (WMAP)}
and {\it Planck} satellites agree with the predictions of the
$\Lambda$CDM cosmology, some anomalous features in the CMB temperature map persist \citep{2013ApJS..208...19H, 2014A&A...566A.135G}.
One of such anomalies is the CMB Cold Spot centred at $(l,b)=(209^\circ, -57^\circ)$ with a temperature decrement $\Delta T\sim-100\mu$K surrounded by a hot ring \citep{2004ApJ...609...22V, 2005MNRAS.356...29C, 2014A&A...571A..23P, 2016A&A...594A..16P}. 
Although a seemingly peculiar 'cold' spot within $\sim7^\circ$ from the center is not significant, 
a structure of the 'cold' spot surrounded by a hot ring within $\sim20^\circ$ is significant at $\gtrsim3\sigma$ assuming that the CMB temperature fluctuation is isotropic Gaussian \citep{2005MNRAS.356...29C, 2010APh....33...69Z, IST10, 2014PhRvD..90j3510N}.

The anomalous feature can be understood through the effects either on the last scattering surface, or large-scale structures between the last scattering surface and us. 
Some inflation and cosmic texture models could imprint such a feature on the last scattering surface \citep{2007Sci...318.1612C, 2008MNRAS.390..913C, 2009PhRvL.103g1301B, 2011JCAP...01..019A}.
On the other hand, large-scale structures can also generate it. 
The anomaly can be explained by the linear/non-linear integrated Sachs-Wolfe (ISW) effect or the Rees-Sciama (RS) effect
\citep{1968Natur.217..511R} caused by a supervoid with a radius of $200-300h^{-1}$Mpc in the 
local Universe \citep{2006ApJ...648...23I,2007ApJ...664..650I,Sakai08,Tomita08, IST10} or  multiple voids
\citep{2016MNRAS.459L..71N}. However, the probability of having such a
large void is low in the $\Lambda$CDM model and multiple voids cannot explain the feature of a hot ring around the Cold Spot.

Motivated by the theoretical predictions, observational searches for a supervoid in the distribution of galaxies began.  
Spectroscopic surveys and observations with photo-z measurements ruled out a supervoid in the redshift range of $0.35<z<1$ \citep{2010MNRAS.404L..69B, 2010ApJ...714..825G}.
At the low redshift $z\sim0.2$, a supervoid was found using the combined
data from the ${\it Pan-STARRS}$ survey and the ${\it WISE-}2{\it MASS}$ galaxy catalog. 
The radius is $R=220h^{-1}$Mpc and the density contrast is $\delta
\sim -0.14$ assuming a top-hat type density profile \citep{2015MNRAS.450..288S, 2016MNRAS.455.1246F}.
\citet{2016MNRAS.462.1882K} also analysed the background galaxies
towards the Cold Spot with the 2{\it MPZ} photometric and 6{\it dF}
spectroscopic redshift measurements and found a low density region in the local Universe. 
However, a void with the observed parameters cannot explain the decrement in the Cold Spot even if large uncertainties in photo-z are included in the observational results. 
\citet{2017arXiv170403814M} observed galaxies spectroscopically over the inner $5^\circ$ of the Cold Spot.
Although the discovered local voids are too small to produce the whole non-Gaussian feature in the Cold Spot via the integrated Sachs-Wolfe effect (ISW), they may indicate some new physics beyond the standard inflationary paradigm.

In fact, the ISW effect from a supervoid towards the Cold Spot is expected to be much weaker than the ordinary Sachs-Wolfe effect from the last scattering surface. 
The feature can be explained by a dense region surrounded by an underdense region at the last scattering surface plus a 'moderately' non-Gaussian supervoid (a density contrast $\delta\sim -0.1$) in the sight line to the Cold Spot \citep{2012MNRAS.421.2731I}. 
Alternatively, it can be explained solely by a non-Gaussian feature imprinted at the last scattering surface. 
Thus to determine the origin of the anomaly, we need to measure the three dimensional matter distribution over the inner $30^\circ$ of the Cold Spot more precisely. 

To do so, we propose to use gravitational lensing, which can directly
measure the total matter (dark + baryon) distribution between a source galaxy and an observer. 
Detecting weak lensing signals from voids with stacked images has been intensively studied \citep{1999MNRAS.309..465A, 2013MNRAS.432.1021H,2013ApJ...762L..20K, 2016MNRAS.459.2762H}. 
Subsequently, observations in the {\it Sloan Digital Sky Survey} and {\it CTIO/Dark Energy Survey} \citep{2012AAS...21941305D} detected such lensing signals \citep{2014MNRAS.440.2922M, 2015MNRAS.454.3357C, 2017MNRAS.465..746S}.
However, these observations can measure only the average density profile of voids. 
It is difficult to detect lensing signals from a single void as the amplitude of density contrast is too small.
While a lensing signal from one void with a radius of $R\sim 10h^{-1}$Mpc is quite weak, signals from a supervoid with a radius of $R\sim 200h^{-1}$Mpc might be possible due to its large tidal shear. 
Although spectroscopic surveys can measure the baryonic matter distribution around the Cold Spot,
the galaxy bias may cause a systematic error in estimating the dark
matter contribution and the cost of conducting a wide field search is relatively high.
In contrast, weak gravitational lensing can directly measure the total
matter distribution with much less model assumptions and cost.

In this paper, we investigate the observational feasibility of finding
supervoids with weak lensing in on-going and future large-scale surveys
such as {\it Subaru/Hyper Suprime-Cam} \citep{2006SPIE.6269E...9M} and DES. 
To assess observational systematics, we used full-sky ray-tracing simulations assuming a
concordant $\Lambda$CDM model. This paper is organised as follows. 
In Section~\ref{sec.data}, we describe the simulation and the analysis methods.
In Section~\ref{sec.result}, we show the result of the lensing analysis. 
We summarize our results in Section~\ref{sec.con},
Throughout this paper, the cosmological parameters follow the WMAP-9yr result \citep{2013ApJS..208...19H}: 
the Hubble parameter $H_0=70$ km/s/Mpc, density parameter of total
matter $\Omega_{\rm m}=0.279$, $\Omega_\Lambda=0.721$, spectral index
$n_s=0.972$ and density fluctuation amplitude $\sigma_8=0.823$.
We assume a flat FLRW cosmology.
Distances and sizes are defined in comoving coordinates if not specified.

\section{Data analysis}
\label{sec.data}
\subsection{basics of weak lensing}
Gravitational lensing is an integrated effect from a matter distribution on a light path. 
In a perturbed FLRW universe, a convergence is described as an integration of the density contrast on a light path \citep{2001PhR...340..291B}
\begin{equation}
\kappa\left(\bm{\theta}\right)=\int^{\infty}_{0} {\rm d}\chi W\left(\chi\right) \delta(\chi\bm{\theta},\chi),
\label{eq.con}
\end{equation}
where $\delta$ is a density contrast at a comoving distance $\chi$ at an angular position $\bm{\theta}$ on a sky-plane.
The lensing weight function $W\left(\chi\right)$ is 
\begin{equation}
W\left(\chi\right)=\frac{3H_0^2\Omega_{\rm m0}}{2c^2}q(\chi)\left\{1+z\left(\chi\right)\right\},
\end{equation}
where $q(\chi)$ is defined as
\begin{equation}
q(\chi)=S_K(\chi) \int^{\infty}_{\chi}{\rm d}\chi' w_{\rm s}\left(\chi'\right) \frac{S_K(\chi'-\chi)}{S_K(\chi')}.
\end{equation}
$w_{\rm s}\left(\chi\right)$ is a number distribution of source galaxies in the line of sight.
$S_K$ is defined as
\begin{equation}
S_K(\chi)=
\begin{cases}
\frac{{\rm sin}\left(\sqrt{K}\chi \right)}{\sqrt{K}} &\hspace{0.5cm}(K>0)\\
\chi &\hspace{0.5cm}(K=0)\\
\frac{{\rm sinh}\left(\sqrt{-K}\chi\right)}{\sqrt{-K}} &\hspace{0.5cm}(K<0).\\
\end{cases}
\end{equation}
$2 \pi S_K(\chi)$ is the circumference of a circle with a radius of $\chi$ in a space with a constant curvature $K$.

The lensing shear is related to the convergence as
\begin{equation}
\gamma\left(\bf{\theta}\right)=\frac{1}{\pi} \int {\rm d}\bm{\theta'} D\left(\bm{\theta}-\bm{\theta'}\right) \kappa\left(\bm{\theta}\right),
\end{equation}
where $D\left(\bm{\theta}-\bm{\theta'}\right)$ is 
\begin{equation}
D\left(\bm{\theta}-\bm{\theta'}\right) = \frac{\theta_2^2-\theta_1^2-2i\theta_1\theta_2}{|\bm{\theta}|^4},
\end{equation}
where $\bm{\theta}=(\theta_1,\theta_2)$.
The average tangential shear at $\theta$ is 
\begin{equation}
\langle\gamma_+\rangle\left(\theta\right) = \langle\kappa\rangle\left(<\theta\right) - \bar{\kappa}\left(\theta\right),
\label{eq.tangen}
\end{equation}
where $\langle\kappa\rangle\left(<\theta\right)$ and $\bar{\kappa}\left(\theta\right)$ are average values of convergence inside a circular aperture and at the edge of the aperture, respectively.
In the weak lensing limit ($|\gamma|\ll1$ and $|\kappa|\ll1$),
tangential shear can be obtained by measuring galaxy ellipticilities. The galaxy shapes are tangentially (radially) aligned if the tangential
shear is positive (negative). The tangential shear takes negative values inside a compensating void \citep{2013MNRAS.432.1021H}.  

Lensing signals are obtained by taking an ensemble average of background galaxy shapes. 
Light emitted from these galaxies traces matter distributions on the light paths. 
Therefore, uncertainties in shapes of background galaxies, and large-scale structures in front and behind a target have to be considered as measurement errors.
In order to estimate the weak lensing detectability of a supervoid taking account of these errors, we define a signal-to-noise ($S/N$) ratio with these uncertainties as
\begin{equation}
\left(\frac{S}{N}\right)^2=\sum_{i,j} \gamma_+(\theta_i) \left[\bm{C}\right]_{ij}^{-1}\gamma_+(\theta_j),
\label{eq.sn}
\end{equation}
where the indices $i$ and $j$ denote the positions of the bins.
$\gamma_+(\theta_i)$ is a tangential shear at $i$-th bin.
The matrix $\bm{C}$ denotes a covariance matrix and $\bm{C}^{-1}$ is the inverse matrix.
In the estimation of a covariance matrix, we assume that the errors come from the above two components.
Then it can be written as
\begin{equation}
\bm{C}=\bm{C}^{\rm shape} + \bm{C}^{\rm LSS}.
\end{equation}
The shape noise is estimated by following \citet{2000MNRAS.313..524V}.
We assume the rms amplitude of the intrinsic elliptisity distribution $\sigma_{\rm e}=0.4$ 
and the background galaxy density $n_{\rm g}=10$ arcmin$^{-2}$, which are typical values for weak lensing observations \citep{2017arXiv170506745M}.
In order to estimate the error coming from large-scale structures, we randomly select 200 positions on sky-planes with 20 realisations.
Then we measure tangential shear values for each selected position and estimate a standard deviation between the profiles.
While we change the number of the profiles for the measurement, the difference in the amplitude of the deviation is $\lesssim1\%$ when we use more than 100 random points.
Therefore, the size of the deviation does not much depend on the number of random points.

\subsection{void model}
\begin{figure}
\begin{center}
\includegraphics[width=0.9\columnwidth]{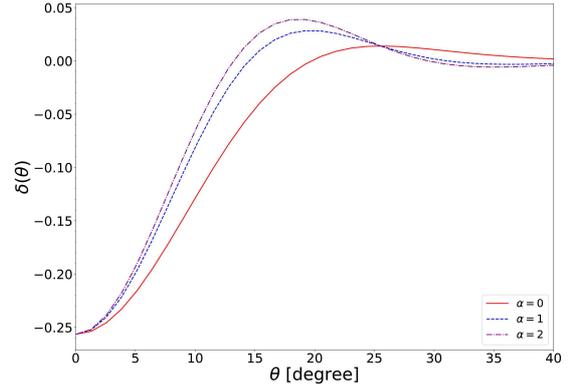}
\caption{Density profile of a void model defined with equation~(\ref{eq.gaussmodel}).
Solid, dashed and dash-dotted lines indicate profiles with $\alpha=0,1$ and $2$, respectively.
$\delta_0=0.29$, $r_0=198h^{-1}$ Mpc and redshift of a void $z_{\rm v}=0.25$ are assumed.}
\label{fig.denprofile}
\end{center}
\end{figure}

In order to model a supervoid, we adopt a compensated Gaussian-like void profile \citep{2016MNRAS.455.1246F}.
We assume that the amplitude of the density contrast is somewhat smaller than 1 and the void radius is sufficiently smaller than the Hubble horizon. 
In this case, the metric can be treated as a local perturbation in the Friedmann-Lema{\^i}tre-Robertson-Walker spacetime. 
Then the density profile of the void is described with a parameter $\alpha$, which characterises the slope of a density profile of a void, as
\begin{equation}
\delta(a,r)=-\delta_0g(a)\left(1-\frac{2+7\alpha}{3+3\alpha}\frac{r^2}{r_0^2}+\frac{2\alpha}{3+3\alpha}\frac{r^4}{r_0^4}\right){\rm exp}\left[-\frac{r^2}{r_0^2}\right],
\label{eq.gaussmodel}
\end{equation}
where $\delta_0$ is a density contrast at the void 
centre at present and $g(a)$ is a growth factor normalised by the present value. 
Figure~\ref{fig.denprofile} shows density profiles for the cases of $\alpha=0, 1$ and 2, respectively.
The adopted parameters are $\delta_0=0.29$, $r_0=198h^{-1}$ Mpc and the redshift of a void $z_{\rm v}=0.25$. 

\subsection{simulation}
In order to investigate the properties of supervoids, we used the 
publicly available data of full-sky ray-tracing simulations\footnote[1]{\url{http://cosmo.phys.hirosaki-u.ac.jp/takahasi/allsky_raytracing/}}.
The details of the simulations are described in Section~2 in \citet{2017arXiv170601472T}.
To carry out $N$-body simulations, they ran the parallel Tree-Particle Mesh code, Gadget2 \citep{2005Natur.435..629S} with $2048^3$ particles.
To generate lensing maps with different source redshifts, they ran
simulations with 14 different box sizes between $450$$h^{-1}$Mpc and $6300$$h^{-1}$Mpc with steps of $450h^{-1}$Mpc. 
There are 6 simulations with different initial conditions for each box size, which can cover
the past light cone of a present observer. 
In order to generate the initial conditions, 
the linear matter transfer function was calculated using {\it CAMB} \citep{2000ApJ...538..473L}.
 
Ray-tracing simulations were carried out by \citet{2001MNRAS.327..169H, 2015MNRAS.453.3043S}.
The light-ray path and the magnification on lens planes were calculated with the multiple-lens plane algorithm.
In order to increase the number of realisations, they chose 18 different positions in each simulation box with the periodic boundary condition.
They used the $N$-body simulations to generate projected mass density shells with a width of $150$$h^{-1}$Mpc.
The lensing information was calculated for 38 different source redshifts between 0 and 5.3.
In order to map lensing signals on a sky-plane, {\it HEALPIX}
\citep{2005ApJ...622..759G} was used to create lensing maps with three
different resolutions, $N_{\rm side}=4096$, $8192$ and $16384$.

In each output of the $N$-body simulations, 
they created dark matter halo catalogues using a software {\it ROCKSTAR} \citep{2013ApJ...762..109B}, which uses the information in the phase space. 
A halo is defined to be a group of more than 50 particles. The adopted
halo masses are $M\geq4\times10^{10}h^{-1}$M$_\odot$. An algorithm 
in {\it HEALPIX} was also used for defining the positions of haloes on the celestial sphere.

Since we focus on the lensing properties of supervoids on large scales, small scale 
structures residing in each void are not important for our analysis. 
Therefore, we used simulations with an angular resolution parameter $N_{\rm side}=4096$, which corresponds to the resolution $\Delta\theta=\sqrt{4\pi/(12N_{\rm side}^2)}=0.85$ arcmin. 
The amplitude of lensing signals depends on the redshifts of a source galaxy and a lens object.
Since the purpose of this paper is to investigate the detectability of
weak lensing signals in the on-going and near-future weak lensing surveys, 
we chose a lens map with a source redshift $z_{\rm s}=0.508$, which is a
typical redshift value for such wide surveys, from the discrete source
redshift planes in the simulations. We used a virial mass as the
definition of the mass of a halo.

\subsection{void finding argorithm}
In order to find voids with haloes, we ran the publicly available VoidFinder code (see details in \citealt{2009ApJ...699.1252F, 2013MNRAS.432.1021H}).
In the VoidFinder, the space is divided into cells and then 
a sphere is enlarged with respect to a cell until the edge of the sphere reaches the third nearest halo.
The spheres are sorted by the radii in decreasing order and marginalized
with larger spheres with a certain criterion.

In our analysis, we used haloes with a mass $M_{\rm vir}\geq2\times10^{14}h^{-1}{\rm M}_\odot$ in the redshift of $0\leq z\leq0.55$.
The processes of running the VoidFinder are quite time consuming, especially for the cases in which low mass haloes are used for full-sky simulations.
Since smoothing convergence maps use only two dimensional information, the computation time can be significantly reduced.  
As a first step, therefore, we look for underdence regions by smoothing convergence maps.
Then we ran the VoidFinder in a part of the regions in each simulation. 
When running the VoidFinder, we used only haloes within $120\times120$ degree$^2$ on a sky-plane which is centred on a smoothed convergence map (see Section~\ref{sec.result}).
The adjustable dimensionless parameter and size parameter in the
VoidFinder are chosen as $\lambda=2.0$ and $\xi=20h^{-1}$Mpc, which are
defined by equation~(5) in \citet{2009ApJ...699.1252F}. We note that the number
of relatively large voids does not much depend on these parameter values (see in \citealt{2013MNRAS.432.1021H}).
In order to select legitimate voids, we exclude voids whose distance
between the void centre and the edges of a selected region is less than the void radius. 

\section{Result}
\label{sec.result}

\begin{figure*}
\subfigure{\includegraphics[width=0.85\columnwidth,bb=0 0 711 769]{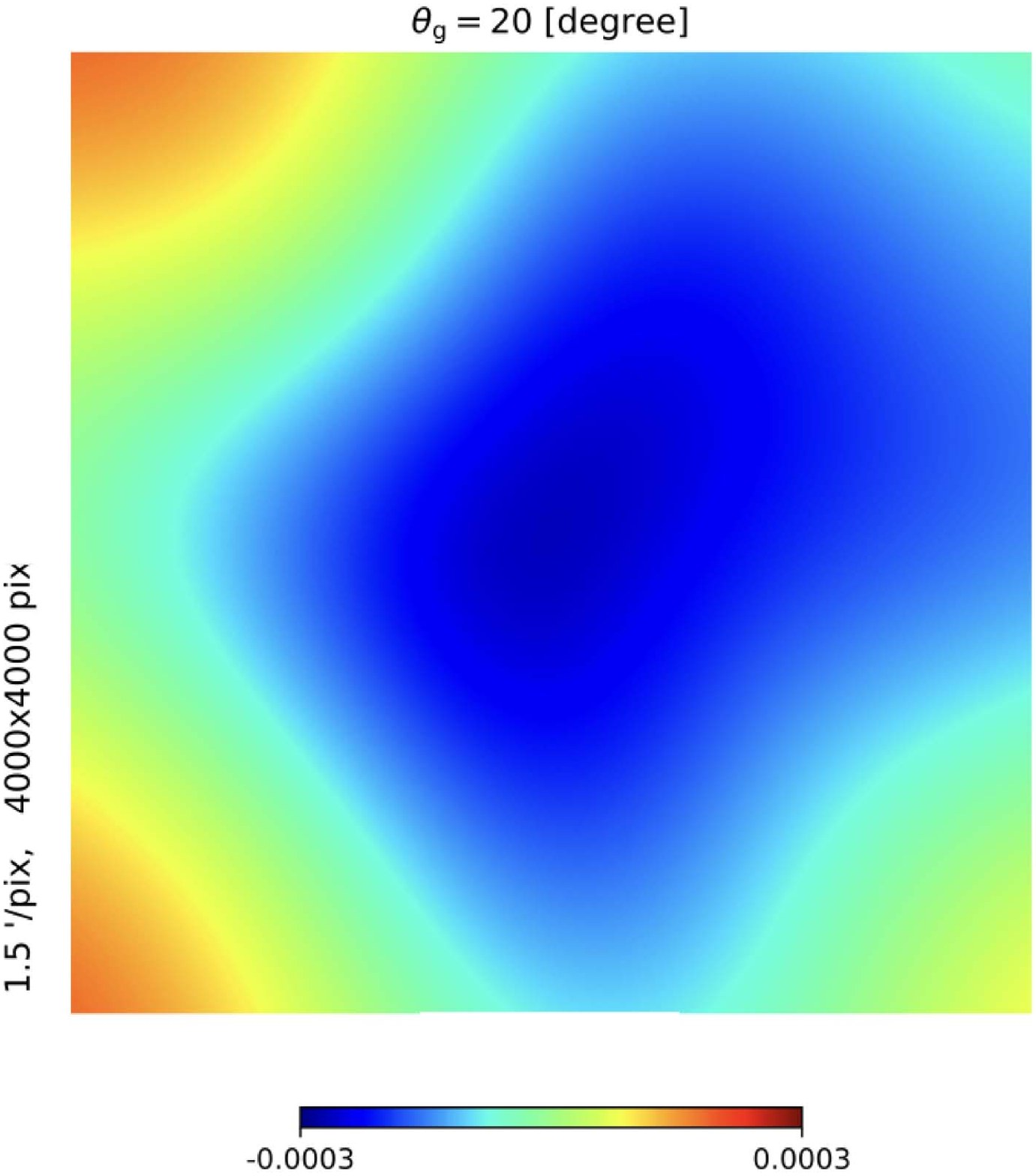}}
\subfigure{\includegraphics[width=1.0\columnwidth,bb=0 0 720 579]{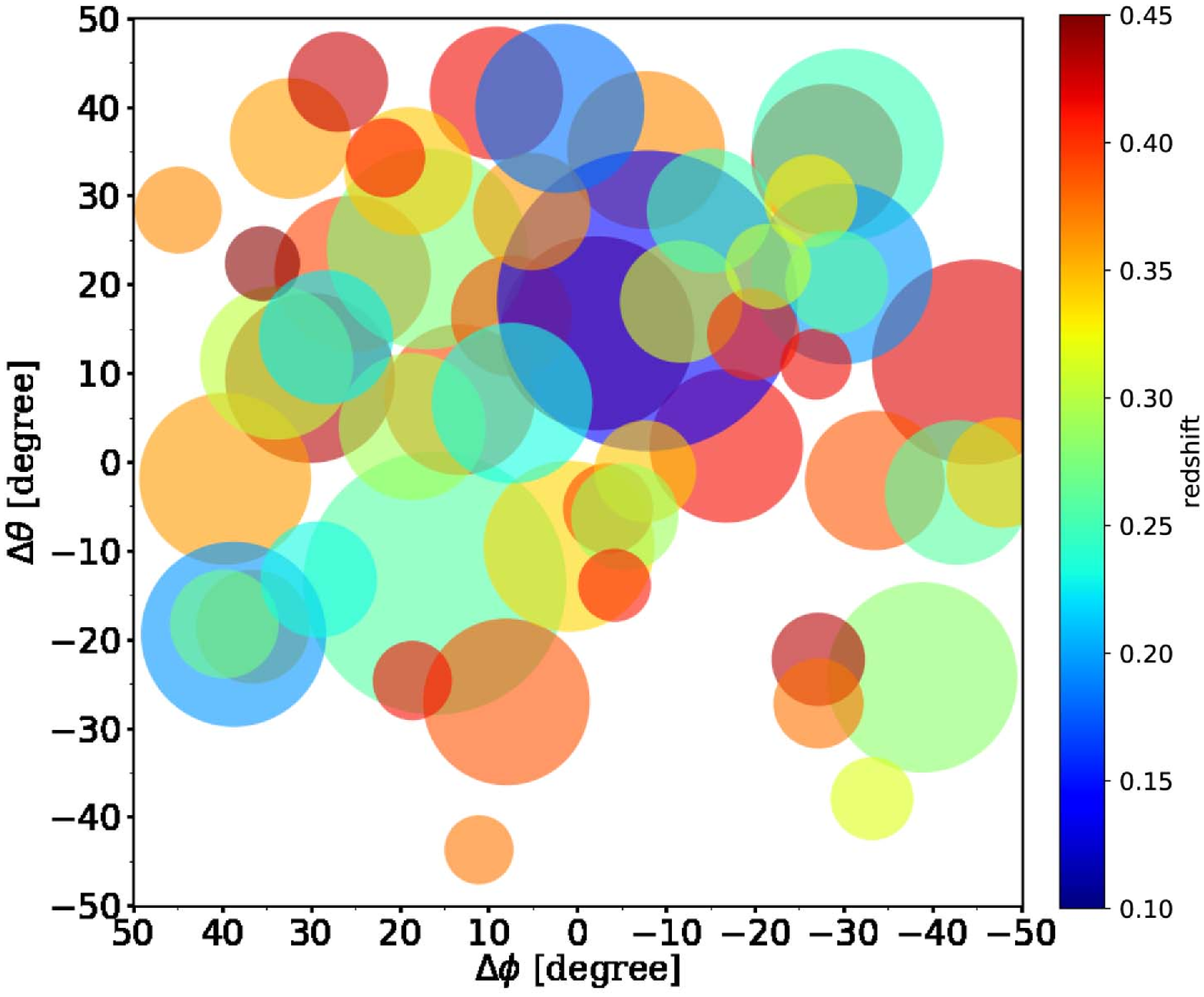}}
\caption{{\it Left:} The convergence map smoothed at a scale of $\theta_{\rm g}=20$ degree. The source redshift is $z_{\rm s}=0.508$ and the field of view is $100\times100$ degree$^2$.
{\it Right:} The positions of voids with respect to the negative
 convergence peak in the redshift range of $0.1\leq z\leq0.45$.
The voids with a radius $R \geq 90h^{-1}$ Mpc are shown.
The horizontal and vertical axes show the coordinates in the simulation. 
The size of a circle shows the angular radius of a void on the sky-plane. 
The colours indicate the redshifts estimated by the VoidFinder. 
The centre of the figure corresponds to the convergence peak in the left panel.}
\label{fig.conv_void}
\end{figure*}

Since running the VoidFinder for all of the full-sky simulations is time consuming, 
we used two dimensional convergence maps for finding a most prominent underdense region whose apparent angular size is similar to that of the Cold Spot.
We used full-sky map lensing simulations with 6 different initial conditions at 18 different observational points, which correspond to $6\times18=108$ realizations.
We used full-sky convergence maps without shape noise and we smoothed the maps with a Gaussian filter with a smoothing scale of  $\theta_g={\rm FWHM}/2\sqrt{2{\rm ln}2}=20$ degrees.
This is the same value used in \citet{2016MNRAS.455.1246F}.
The left panel in Figure~\ref{fig.conv_void} shows a smoothed convergence map centred at the largest negative convergence peak.
The size of the figure is $100\times100$ degree$^2$ with a pixel scale of $1.5$ arcmin. 
Finding the position of a maximum of a potential in a large underdense region is not an easy task in real observations.
The size of the smoothing scale would shift the peak of the potential and weaken the amplitude of the tangential shear.
However, the position of the largest negative convergence peak is not much affected by
the smoothing scale if larger than several tens of degrees.

In order to measure the properties of voids, we ran 
the VoidFinder by selecting haloes within $\sim15,000$ degree$^2$ around
the largest negative peak in the smoothed convergence map.
We used haloes with a mass of $M\geq2\times10^{14}h^{-1}{\rm M}_\odot$ in the redshift range of $0\leq z\leq0.55$.
The average distance between haloes is $\sim60h^{-1}$Mpc.
The right panel in Figure~\ref{fig.conv_void} shows the positions of
voids projected onto the sky-plane. The sizes of the circles and colours
represent the sizes of the voids on the sky-plane and their redshifts, respectively.
The centre of the figure corresponds to the convergence peak.
The void residing at $(\Delta\phi, \Delta\theta)\sim(-2^\circ, 15^\circ)$ is the
largest one at $z=0.438$ with a radius $r=228~h^{-1}$Mpc.
As shown in Figure~\ref{fig.conv_void}, smaller voids are clustering in front and behind the largest void.
In order to estimate the rarity of the underdense region that consist of several tens of voids, 
we use an 'occupation ratio' defined as a ratio of the total volume occupied by voids to the cosmic volume within a certain angular radius in the redshift range of $0.2 \leq z \leq 0.5$. 
To calculate the average occupation ratios, we selected haloes with a mass of $M\geq2\times10^{14}h^{-1}{\rm M}_\odot$ from 10 realisations and ran 
the VoidFinder in a half region of a hemisphere in each realisation.
We checked that the number of realisations does not largely affect the significance.
We used the Monte Carlo method to calculate the volumes of voids. 
When the centre of a void falls within a certain angular distance from 
the position of the largest negative convergence peak, the volume of the
void is added to the one occupied by voids. 
The average and the standard deviation of the occupation ratio were
obtained from values centred at randomly selected 3 points on a
sky-plane for each realisation, i.e, the total number of random points are $3\times10=30$ for 10 realisations.
Table~\ref{tab.aven} shows the results for 5 different choices of the
radial distance $r$.
We found that the occupation ratio around the largest negative convergence peak shows an excess for $r=25^\circ - 30^\circ$. 
Although the largest void occupies the most volume at the relevant redshift range, 
the excess cannot be solely explained by the largest void itself because the angular radius is just $\sim 15$ degrees.
Voids in {\it Void-in-void} mode are preferentially clustered \citep{2004MNRAS.350..517S,2013MNRAS.434.1435C,2017MNRAS.468.4822L}. 
Our result is consistent with the previous results, and the observed large dip in the smoothed convegence map was created by both the largest and surrounding smaller voids. 

\begin{table}
\caption{The void occupation ratios in the vicinity of the largest negative convergence peak in the redshift range of 0.2-0.5.
They are estimated from the volumes of voids whose centre is within a certain radius from the peak. 
The VoidFinder is run in a half region of a hemisphere for 10 realisations.
The average and the standard deviation were calculated from 30 samples of 'field' points. 
The volumes are estimated by the VoidFinder.
Column (1): angular radius , Column (2): average and standard
 deviation around 30 'field' points, Column (3):
 ratio around the largest negative convergence peak. }
\begin{center}
\begin{tabular}{|c|c|c|}
radius            & 30 'field' points & peak\\ \hline\hline
$r=25^\circ$& $0.312\pm0.024$ & $0.584$\\
$r=28^\circ$& $0.306\pm0.021$ & $0.466$\\
$r=30^\circ$& $0.312\pm0.019$ & $0.508$\\
$r=33^\circ$& $0.304\pm0.013$ & $0.573$\\
$r=35^\circ$& $0.305\pm0.017$ & $0.602$
\end{tabular}
\end{center}
\label{tab.aven}
\end{table}

\begin{figure}
\includegraphics[width=1.\columnwidth]{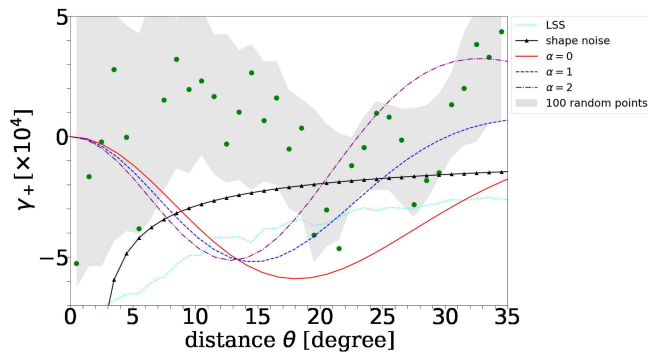}
\caption{The tangential shear profiles for a supervoid.
The horizontal axis shows the distance from the largest negative convergence peak. 
The vertical axis shows the tangential shear. 
Green points show the profile in the simulation.
The grey shaded line shows the average tangential shear profile with respect to 100 randomly selected points within $5$ degrees from the peak. 
The shaded region indicates the $1\sigma$ standard deviations. 
The line with triangles and dotted line (light blue) show the errors from shape noise and large-scale structures, respectively. 
A galaxy number density $n_{\rm g}=10$ arcmin$^{-2}$, an intrinsic ellipticity dispersion $\sigma_{\rm e}=0.4$ and a source redshift $z_{\rm s}=0.508$ are assumed. 
Solid, dashed and dash-dotted lines show the tangential profiles derived from equation~(\ref{eq.gaussmodel}) for $\alpha=0,1$ and 2, respectively.  
$\delta_0=0.29$, $r_0=198h^{-1}$Mpc and $z_l=0.22$ which were obtained in \citet{2016MNRAS.455.1246F} are adopted for the model.
We assume that the position of the peak correspond to the centre of the void. }
\label{fig.shearprofile}
\end{figure}

The analytical tangential shear profile for one single supervoid can be 
calculated from equation~(\ref{eq.con}), (\ref{eq.tangen}) and (\ref{eq.gaussmodel}). 
The solid, dashed and dash-dotted lines in Figure~\ref{fig.shearprofile} show the analytical profiles for different $\alpha$ values
by assuming the parameters in \citet{2016MNRAS.455.1246F}.
The line with triangles shows the error from shape noise assuming a galaxy number density $n_{\rm g}=10$ arcmin$^{-2}$.
The dotted line (light blue) shows the errors from large-scale structures.
As shown in Figure~\ref{fig.denprofile}, the shear profiles do not show large differences in the inner region for the three different void profiles.
However, the differences in the steepness of the density profile in the ridge region can generate the differences in the lensing signals in that region.
The green points in Figure~\ref{fig.shearprofile} shows the tangential shear profile in our simulation as a function of distance from the largest negative convergence peak.
As noticed, the position of the peak in the smoothed convergence map is not much affected by the smoothing scale if larger than several tens of degrees. 
In order to estimate the model variability due to ambiguity in the choice of the coordinate centre, however, 
we measured the tangential shear profile for 100 randomly selected centres within 5 degrees from the peak in the smoothed convergence map. 
The grey shaded line in figure 3 shows the average tangential shear
profile at 100 random points with the $1\sigma$ standard deviation.
We divided the angular distance from 0 to 35 degrees into 35 bins, i.e., the bin size $\Delta\theta$ is 1 degree.
The source redshift is set to $z_{\rm s}=0.508$ for the analytical and simulated profiles.
Although we measured the tangential shear profiles with different bin sizes, 
it turned out that the shear profiles do not much depend on the bin size.  
The estimated signal-to-noise ratio for the tangential shear at angular
distances 10 to 35 degrees from the centre is turned out to be $S/N\sim 3$. 
The result indicates that the lensing signals from a locally
underdence region can be marginally detected with weak lensing depending on the scale of mass deficiency.
In our method, the shear is measured as a function of a distance from the centre of a void.
If the shear is measured as a function of a distance from a void boundary, the lensing signal can be enhanced by a factor of two \citep{2016MNRAS.457.2540C}.
Since the amplitude of the average tangential shear at the wall of a void is a measure of the average density contrast inside the wall,
an increase in the amplitude of density contrast at the void centre causes an increase in the amplitude of the average shear, and an increase in the void radius causes an increase in the distance of the negative peak position of the shear measured from the void centre.

\begin{figure*}
\hspace{2cm}
\subfigure{\includegraphics[width=1.5\columnwidth]{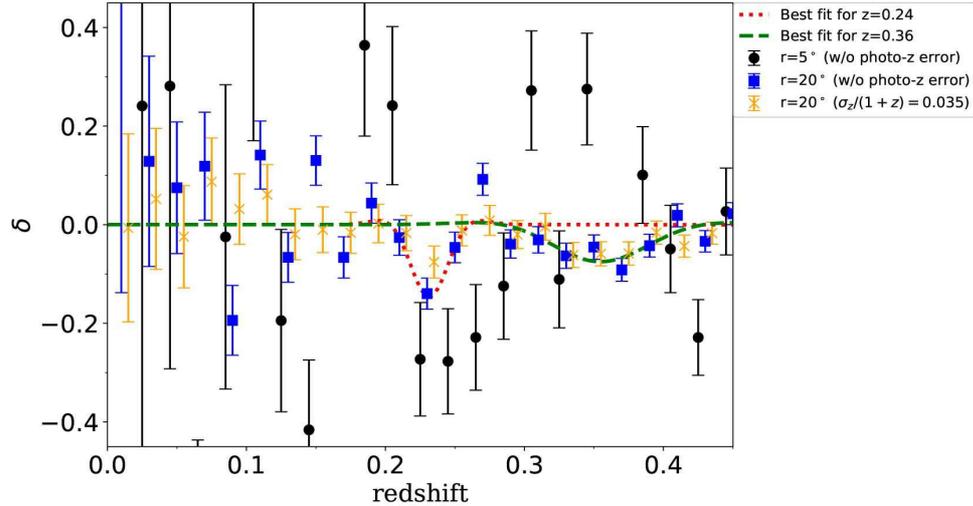}}
\caption{The density contrast towards the largest negative convergence peak as a function of redshift without and with photo-z errors.
The horizontal axis shows the redshift and vertical axis shows the density contrast.
The circles and squares are the density contrast within a radius of $5$ and $20$ degrees without photo-z errors.
The crosses show the density contrast with photo-z errors, respectively.
The results for the third, fifth and eighth bins with a radius of 5 degree are out of the plotted range.
The density contrast was obtained by using haloes within a certain  radius on a sky-plane from the negative convergence peak.  
The red dotted and green dashed lines show the best fitted curves with the result of a radius of 20 degrees for the underdensities exist at z=0.23 and 0.36, respectively.
We put the void model at redshifts of $z=0.25$ and $0.35$ and fitted the data without photo-z errors assuming equation~(\ref{eq.gaussmodel}), respectively.  
$\delta_0$, $r_0$ and z are the free parameters in the fitting.
The error bars show Poissonian noises. 
The standard deviation of the photo-z errors is assumed to be $\sigma_z=0.035(1+z_{\rm halo})$.}
\label{fig.densdis}
\end{figure*}

Figure~\ref{fig.densdis} shows the density contrast as a function of the
redshift (see \citealt{2015MNRAS.450..288S}). 
In our simulation, haloes with a mass of $M\geq3\times 10^{13}h^{-1}$M$_\odot$ were used to measure
the number density at each redshift bin within a certain radius with
respect to the largest negative convergence peak. 
The errors were assumed to be Poissonian. 
In order to reduce the errors, lower mass haloes were used for this analysis.
However, the results with a threshold of $M\geq3\times 10^{13}h^{-1}$M$_\odot$ and $M\geq2\times 10^{14}h^{-1}$M$_\odot$ are consistent with each other within $2\sigma$.
The average number density at each redshift bin was estimated by using the rest of haloes in the simulation.
To investigate the inner density profile of the largest void, we divided the redshift range
$0\leq z\leq0.6$ into 30 bins, i.e., the bin size is $\Delta z=0.02$.
As shown in figure 4, we can see a large dip at $z\sim 0.36$
and a small dip at $z \sim 0.23$ if smoothed within $20^\circ$ from the
negative convergence peak. Since any large underdence regions are
associated with multiple voids, such a clustering of multiple voids in
the line of sight may be consistent with a recent spectroscopic
observation toward the Cold Spot\citep{2017arXiv170403814M}. 

\begin{table*}
\caption{The fitting results with the void model for the density
 profiles at low and high redshifts with a radius of 20 degrees. 
The fitting was carried out by putting the void model at the redshift of $z=0.25$ and $z=0.35$, respectively. 
The errors indicate $1\sigma$.
The photo-z errors are randomly added to the redshifts of haloes
 assuming a Gaussian distribution for the errors.
Column (1): deviation of photo-z error, Column (2)-(4): fitted values (density contrast, radius and redshift) defined in equation~(\ref{eq.gaussmodel}).}
\begin{center}
\begin{tabular}{|c|c|c|c|}
$\sigma_z/(1+z_{\rm halo})$        &     $\delta_0$               & $r_0$ [$h^{-1}$Mpc]   & $z_{l}$\\ \hline\hline
low-z underdensity&&&\\ \hline
0                            &  $0.1632\pm0.0044$      & $63\pm2$             & $0.2315\pm0.0005$\\
0.035                     &  $0.0854\pm0.0014$      & $58\pm1$             & $0.2290\pm0.0003$\\
\hline\hline
high-z underdensity&&&\\
\hline
0                            &  $0.0895\pm0.0024$      & $151\pm4$           & $0.3552\pm0.0013$\\
0.035                     &  $0.0674\pm0.0059$      & $208\pm2$           & $0.3626\pm0.0044$\\
\end{tabular}
\end{center}
\label{tab.fit}
\end{table*}

We added photo-z errors to the redshifts of haloes 
and investigated the differences of parameters by fitting the density profile.
To take into account the photo-z errors, we randomly added 
Gaussian errors to the redshifts of haloes. The error value was assumed
to be $\sigma_z/(1+z_{\rm halo})=0.035$ as in \citet{2015MNRAS.450..288S}.
Table~\ref{tab.fit} shows the fitting results in the cases with and without photo-z errors.
We fitted the each underdensity which exist at $z\sim0.23$ and $z\sim0.36$ with the void model defined by equation~(\ref{eq.gaussmodel} ), respectively.
The amplitude of density contrast decreases if photo-z errors were added to the data; Structures such as voids, filaments and walls are smoothed out.
In \citet{2015MNRAS.450..288S}, only the density contrast was used to study the void properties. 
They found that the values of the density contrast measured with smaller radius are higher than that with a larger radius due to the averaging process.
Recently, \citet{2017arXiv170403814M} analysed the data with spectroscopic samples within the inner $5$ degrees of the Cold Spot.  
They found a few voids with a radius of $50-170h^{-1}$Mpc in the redshift range of $0\leq z\leq0.5$.
Although the numbers of their samples and the area are small,  their results seem to be consistent with our simulation result.
The errors in photo-z mitigate small substructures in supervoids which can be seen in spectroscopic observations.
Moreover, reconstruction of three dimensional mass distributions from weak lensing information is not an easy task \citep{2001astro.ph.11605T,2002PhRvD..66f3506H,2003MNRAS.344.1307B}.
A combination of photometric and spectroscopic observations enable to measure precise matter distribution toward supervoids.

Assuming the best-fitted supervoid model in \citet{2016MNRAS.455.1246F},
it is effective to observe an annulus region around $\theta=20$ degrees for detecting the lensing signals 
since the tangential shear of an underdence region is maximized at $\theta=20$ degrees on a sky-plane with respect to the centre of the void.
Compared with the fitting results, in addition, the measured density
contrast in \citet{2016MNRAS.455.1246F} is $1 - 5$ times larger than that in our simulation. 
Since the tangential shear is proportional to the density contrast, and the amplitude of it depends on the redshifts of source galaxies and a supervoid, 
the lensing signals can be detected at $S/N\gtrsim4$ at an angular distance between 10 and 35 degrees in on-going and future
weak lensing observations.

\section{Conclusions}
\label{sec.con}
We have studied the expected weak lensing signal from a possible large underdence region towards the Cold Spot.
In order to do so, we have used all-sky ray-tracing simulations in a
concordant $\Lambda$CDM model consistent with the WMAP 9yr results. In
order to find a most prominent underdense region that can contribute to 
the Cold Spot via the integrated Sachs-Wolfe effect, 
we smoothed the obtained convergence maps at a scale of $\sim 20$ degrees.
Then we ran the VoidFinder in the region for investigating the properties and measured the tangential shear profile around the largest negative peak in the convergence maps. 
In our simulation, the signal of the negative tangential shear in the locally underdense region expected in a concordant $\Lambda$CDM model can be detected at $S/N\sim3$. We found that a number of voids is
clustering around the negative convergence peak.  
If a single supervoid with a radius of $\sim 200\,h^{-1}\,\textrm{Mpc}$ and a density contrast $\delta\sim -0.3$ resides at a redshift $z\sim 0.2$ as
observationally suggested in \citet{2016MNRAS.455.1246F}, the lensing signal can  be detected at $S/N\sim4-10$ without resorting to stacking.

Weak lensing measurements can test the existence of a large underdence region toward the Cold Spot
and the shape of the tangential shear profile can give a direct measurement  for the steepness of the matter density in the ridge region of the supervoid.
By combining weak lensing measurements, on-going surveys such as {\it Subaru/Primary Spectroscopic Survey}
\citep{2012SPIE.8446E..0YS, 2014SPIE.9147E..0TS, 2016SPIE.9908E..1MT}
would certainly benefit us for investigating a three dimensional matter distribution in the local Universe to the Cold Spot.

\section*{Acknowledgements}
We thank an anonymous referee for giving useful comments and improving the manuscript. 
We thank C. Foster for the use of the VoidFinder.
We also thank R. Takahashi, T. Hamana and M. Shirasaki for carrying out the simulations.
We would like to thank K. Umetsu, I. Chiu, T. Okumura, Y.Toba for useful comments and discussions.
YH is supported by ASIAA, Taiwan. 
Numerical computations presented in this paper were in part carried out on the general-purpose PC farm at Center for Computational Astrophysics, CfCA, of National Astronomical Observatory of Japan.
Data analyses were (in part) carried out on common use data analysis computer system at the Astronomy Data Center, ADC, of the National Astronomical Observatory of Japan.





\bibliographystyle{mnras}
\bibliography{mn-jour,bibtex}

\begin{thebibliography}{}
\makeatletter
\relax
\def\mn@urlcharsother{\let\do\@makeother \do\$\do\&\do\#\do\^\do\_\do\%\do\~}
\def\mn@doi{\begingroup\mn@urlcharsother \@ifnextchar [ {\mn@doi@}
  {\mn@doi@[]}}
\def\mn@doi@[#1]#2{\def\@tempa{#1}\ifx\@tempa\@empty \href
  {http://dx.doi.org/#2} {doi:#2}\else \href {http://dx.doi.org/#2} {#1}\fi
  \endgroup}
\def\mn@eprint#1#2{\mn@eprint@#1:#2::\@nil}
\def\mn@eprint@arXiv#1{\href {http://arxiv.org/abs/#1} {{\tt arXiv:#1}}}
\def\mn@eprint@dblp#1{\href {http://dblp.uni-trier.de/rec/bibtex/#1.xml}
  {dblp:#1}}
\def\mn@eprint@#1:#2:#3:#4\@nil{\def\@tempa {#1}\def\@tempb {#2}\def\@tempc
  {#3}\ifx \@tempc \@empty \let \@tempc \@tempb \let \@tempb \@tempa \fi \ifx
  \@tempb \@empty \def\@tempb {arXiv}\fi \@ifundefined
  {mn@eprint@\@tempb}{\@tempb:\@tempc}{\expandafter \expandafter \csname
  mn@eprint@\@tempb\endcsname \expandafter{\@tempc}}}

\bibitem[\protect\citeauthoryear{{Afshordi}, {Slosar}  \& {Wang}}{{Afshordi}
  et~al.}{2011}]{2011JCAP...01..019A}
{Afshordi} N.,  {Slosar} A.,   {Wang} Y.,  2011, \mn@doi [JCAP]
  {10.1088/1475-7516/2011/01/019}, \href
  {http://adsabs.harvard.edu/abs/2011JCAP...01..019A} {1, 019}

\bibitem[\protect\citeauthoryear{{Amendola}, {Frieman}  \& {Waga}}{{Amendola}
  et~al.}{1999}]{1999MNRAS.309..465A}
{Amendola} L.,  {Frieman} J.~A.,   {Waga} I.,  1999, \mn@doi [MNRAS]
  {10.1046/j.1365-8711.1999.02841.x}, \href
  {http://adsabs.harvard.edu/abs/1999MNRAS.309..465A} {309, 465}

\bibitem[\protect\citeauthoryear{{Bacon} \& {Taylor}}{{Bacon} \&
  {Taylor}}{2003}]{2003MNRAS.344.1307B}
{Bacon} D.~J.,  {Taylor} A.~N.,  2003, \mn@doi [MNRAS]
  {10.1046/j.1365-8711.2003.06922.x}, \href
  {http://adsabs.harvard.edu/abs/2003MNRAS.344.1307B} {344, 1307}

\bibitem[\protect\citeauthoryear{{Bartelmann} \& {Schneider}}{{Bartelmann} \&
  {Schneider}}{2001}]{2001PhR...340..291B}
{Bartelmann} M.,  {Schneider} P.,  2001, \mn@doi [Physical Rep.]
  {10.1016/S0370-1573(00)00082-X}, \href
  {http://adsabs.harvard.edu/abs/2001PhR...340..291B} {340, 291}

\bibitem[\protect\citeauthoryear{{Behroozi}, {Wechsler}  \& {Wu}}{{Behroozi}
  et~al.}{2013}]{2013ApJ...762..109B}
{Behroozi} P.~S.,  {Wechsler} R.~H.,   {Wu} H.-Y.,  2013, \mn@doi [ApJ]
  {10.1088/0004-637X/762/2/109}, \href
  {http://adsabs.harvard.edu/abs/2013ApJ...762..109B} {762, 109}

\bibitem[\protect\citeauthoryear{{Bond}, {Frolov}, {Huang}  \& {Kofman}}{{Bond}
  et~al.}{2009}]{2009PhRvL.103g1301B}
{Bond} J.~R.,  {Frolov} A.~V.,  {Huang} Z.,   {Kofman} L.,  2009, \mn@doi
  [Phys. Rev. Let.] {10.1103/PhysRevLett.103.071301}, \href
  {http://adsabs.harvard.edu/abs/2009PhRvL.103g1301B} {103, 071301}

\bibitem[\protect\citeauthoryear{{Bremer}, {Silk}, {Davies}  \&
  {Lehnert}}{{Bremer} et~al.}{2010}]{2010MNRAS.404L..69B}
{Bremer} M.~N.,  {Silk} J.,  {Davies} L.~J.~M.,   {Lehnert} M.~D.,  2010,
  \mn@doi [MNRAS] {10.1111/j.1745-3933.2010.00837.x}, \href
  {http://adsabs.harvard.edu/abs/2010MNRAS.404L..69B} {404, L69}

\bibitem[\protect\citeauthoryear{{Cautun}, {Cai}  \& {Frenk}}{{Cautun}
  et~al.}{2016}]{2016MNRAS.457.2540C}
{Cautun} M.,  {Cai} Y.-C.,   {Frenk} C.~S.,  2016, \mn@doi [MNRAS]
  {10.1093/mnras/stw154}, \href
  {http://adsabs.harvard.edu/abs/2016MNRAS.457.2540C} {457, 2540}

\bibitem[\protect\citeauthoryear{{Ceccarelli}, {Paz}, {Lares}, {Padilla}  \&
  {Lambas}}{{Ceccarelli} et~al.}{2013}]{2013MNRAS.434.1435C}
{Ceccarelli} L.,  {Paz} D.,  {Lares} M.,  {Padilla} N.,   {Lambas} D.~G.,
  2013, \mn@doi [MNRAS] {10.1093/mnras/stt1097}, \href
  {http://adsabs.harvard.edu/abs/2013MNRAS.434.1435C} {434, 1435}

\bibitem[\protect\citeauthoryear{{Clampitt} \& {Jain}}{{Clampitt} \&
  {Jain}}{2015}]{2015MNRAS.454.3357C}
{Clampitt} J.,  {Jain} B.,  2015, \mn@doi [MNRAS] {10.1093/mnras/stv2215},
  \href {http://adsabs.harvard.edu/abs/2015MNRAS.454.3357C} {454, 3357}

\bibitem[\protect\citeauthoryear{{Cruz}, {Mart{\'{\i}}nez-Gonz{\'a}lez},
  {Vielva}  \& {Cay{\'o}n}}{{Cruz} et~al.}{2005}]{2005MNRAS.356...29C}
{Cruz} M.,  {Mart{\'{\i}}nez-Gonz{\'a}lez} E.,  {Vielva} P.,   {Cay{\'o}n} L.,
  2005, \mn@doi [MNRAS] {10.1111/j.1365-2966.2004.08419.x}, \href
  {http://adsabs.harvard.edu/abs/2005MNRAS.356...29C} {356, 29}

\bibitem[\protect\citeauthoryear{{Cruz}, {Turok}, {Vielva},
  {Mart{\'{\i}}nez-Gonz{\'a}lez}  \& {Hobson}}{{Cruz}
  et~al.}{2007}]{2007Sci...318.1612C}
{Cruz} M.,  {Turok} N.,  {Vielva} P.,  {Mart{\'{\i}}nez-Gonz{\'a}lez} E.,
  {Hobson} M.,  2007, \mn@doi [Science] {10.1126/science.1148694}, \href
  {http://adsabs.harvard.edu/abs/2007Sci...318.1612C} {318, 1612}

\bibitem[\protect\citeauthoryear{{Cruz}, {Mart{\'{\i}}nez-Gonz{\'a}lez},
  {Vielva}, {Diego}, {Hobson}  \& {Turok}}{{Cruz}
  et~al.}{2008}]{2008MNRAS.390..913C}
{Cruz} M.,  {Mart{\'{\i}}nez-Gonz{\'a}lez} E.,  {Vielva} P.,  {Diego} J.~M.,
  {Hobson} M.,   {Turok} N.,  2008, \mn@doi [MNRAS]
  {10.1111/j.1365-2966.2008.13812.x}, \href
  {http://adsabs.harvard.edu/abs/2008MNRAS.390..913C} {390, 913}

\bibitem[\protect\citeauthoryear{{Diehl} \& {Dark Energy Survey
  Collaboration}}{{Diehl} \& {Dark Energy Survey
  Collaboration}}{2012}]{2012AAS...21941305D}
{Diehl} H.~T.,  {Dark Energy Survey Collaboration} 2012, in American
  Astronomical Society Meeting Abstracts \#219. p. 413.05

\bibitem[\protect\citeauthoryear{{Finelli}, {Garc{\'{\i}}a-Bellido},
  {Kov{\'a}cs}, {Paci}  \& {Szapudi}}{{Finelli}
  et~al.}{2016}]{2016MNRAS.455.1246F}
{Finelli} F.,  {Garc{\'{\i}}a-Bellido} J.,  {Kov{\'a}cs} A.,  {Paci} F.,
  {Szapudi} I.,  2016, \mn@doi [MNRAS] {10.1093/mnras/stv2388}, \href
  {http://adsabs.harvard.edu/abs/2016MNRAS.455.1246F} {455, 1246}

\bibitem[\protect\citeauthoryear{{Foster} \& {Nelson}}{{Foster} \&
  {Nelson}}{2009}]{2009ApJ...699.1252F}
{Foster} C.,  {Nelson} L.~A.,  2009, \mn@doi [ApJ]
  {10.1088/0004-637X/699/2/1252}, \href
  {http://adsabs.harvard.edu/abs/2009ApJ...699.1252F} {699, 1252}

\bibitem[\protect\citeauthoryear{{G{\'o}rski}, {Hivon}, {Banday}, {Wandelt},
  {Hansen}, {Reinecke}  \& {Bartelmann}}{{G{\'o}rski}
  et~al.}{2005}]{2005ApJ...622..759G}
{G{\'o}rski} K.~M.,  {Hivon} E.,  {Banday} A.~J.,  {Wandelt} B.~D.,  {Hansen}
  F.~K.,  {Reinecke} M.,   {Bartelmann} M.,  2005, \mn@doi [ApJ]
  {10.1086/427976}, \href {http://adsabs.harvard.edu/abs/2005ApJ...622..759G}
  {622, 759}

\bibitem[\protect\citeauthoryear{{Granett}, {Szapudi}  \& {Neyrinck}}{{Granett}
  et~al.}{2010}]{2010ApJ...714..825G}
{Granett} B.~R.,  {Szapudi} I.,   {Neyrinck} M.~C.,  2010, \mn@doi [ApJ]
  {10.1088/0004-637X/714/1/825}, \href
  {http://adsabs.harvard.edu/abs/2010ApJ...714..825G} {714, 825}

\bibitem[\protect\citeauthoryear{{Gurzadyan}, {Kashin}, {Khachatryan},
  {Poghosian}, {Sargsyan}  \& {Yegorian}}{{Gurzadyan}
  et~al.}{2014}]{2014A&A...566A.135G}
{Gurzadyan} V.~G.,  {Kashin} A.~L.,  {Khachatryan} H.,  {Poghosian} E.,
  {Sargsyan} S.,   {Yegorian} G.,  2014, \mn@doi [A\&A]
  {10.1051/0004-6361/201423565}, \href
  {http://adsabs.harvard.edu/abs/2014A%26A...566A.135G} {566, A135}

\bibitem[\protect\citeauthoryear{{Hamana} \& {Mellier}}{{Hamana} \&
  {Mellier}}{2001}]{2001MNRAS.327..169H}
{Hamana} T.,  {Mellier} Y.,  2001, \mn@doi [MNRAS]
  {10.1046/j.1365-8711.2001.04685.x}, \href
  {http://adsabs.harvard.edu/abs/2001MNRAS.327..169H} {327, 169}

\bibitem[\protect\citeauthoryear{{Higuchi} \& {Shirasaki}}{{Higuchi} \&
  {Shirasaki}}{2016}]{2016MNRAS.459.2762H}
{Higuchi} Y.,  {Shirasaki} M.,  2016, \mn@doi [MNRAS] {10.1093/mnras/stw814},
  \href {http://adsabs.harvard.edu/abs/2016MNRAS.459.2762H} {459, 2762}

\bibitem[\protect\citeauthoryear{{Higuchi}, {Oguri}  \& {Hamana}}{{Higuchi}
  et~al.}{2013}]{2013MNRAS.432.1021H}
{Higuchi} Y.,  {Oguri} M.,   {Hamana} T.,  2013, \mn@doi [MNRAS]
  {10.1093/mnras/stt521}, \href
  {http://adsabs.harvard.edu/abs/2013MNRAS.432.1021H} {432, 1021}

\bibitem[\protect\citeauthoryear{{Hinshaw} et~al.,}{{Hinshaw}
  et~al.}{2013}]{2013ApJS..208...19H}
{Hinshaw} G.,  et~al., 2013, \mn@doi [ApJS] {10.1088/0067-0049/208/2/19}, \href
  {http://adsabs.harvard.edu/abs/2013ApJS..208...19H} {208, 19}

\bibitem[\protect\citeauthoryear{{Hu} \& {Keeton}}{{Hu} \&
  {Keeton}}{2002}]{2002PhRvD..66f3506H}
{Hu} W.,  {Keeton} C.~R.,  2002, \mn@doi [Phys. Rev. D]
  {10.1103/PhysRevD.66.063506}, \href
  {http://adsabs.harvard.edu/abs/2002PhRvD..66f3506H} {66, 063506}

\bibitem[\protect\citeauthoryear{{Inoue}}{{Inoue}}{2012}]{2012MNRAS.421.2731I}
{Inoue} K.~T.,  2012, \mn@doi [MNRAS] {10.1111/j.1365-2966.2012.20513.x}, \href
  {http://adsabs.harvard.edu/abs/2012MNRAS.421.2731I} {421, 2731}

\bibitem[\protect\citeauthoryear{{Inoue} \& {Silk}}{{Inoue} \&
  {Silk}}{2006}]{2006ApJ...648...23I}
{Inoue} K.~T.,  {Silk} J.,  2006, \mn@doi [ApJ] {10.1086/505636}, \href
  {http://adsabs.harvard.edu/abs/2006ApJ...648...23I} {648, 23}

\bibitem[\protect\citeauthoryear{{Inoue} \& {Silk}}{{Inoue} \&
  {Silk}}{2007}]{2007ApJ...664..650I}
{Inoue} K.~T.,  {Silk} J.,  2007, \mn@doi [ApJ] {10.1086/517603}, \href
  {http://adsabs.harvard.edu/abs/2007ApJ...664..650I} {664, 650}

\bibitem[\protect\citeauthoryear{{Inoue}, {Sakai}  \& {Tomita}}{{Inoue}
  et~al.}{2010}]{IST10}
{Inoue} K.~T.,  {Sakai} N.,   {Tomita} K.,  2010, \mn@doi [ApJ]
  {10.1088/0004-637X/724/1/12}, \href
  {http://adsabs.harvard.edu/abs/2010ApJ...724...12I} {724, 12}

\bibitem[\protect\citeauthoryear{{Kov{\'a}cs} \&
  {Garc{\'{\i}}a-Bellido}}{{Kov{\'a}cs} \&
  {Garc{\'{\i}}a-Bellido}}{2016}]{2016MNRAS.462.1882K}
{Kov{\'a}cs} A.,  {Garc{\'{\i}}a-Bellido} J.,  2016, \mn@doi [MNRAS]
  {10.1093/mnras/stw1752}, \href
  {http://adsabs.harvard.edu/abs/2016MNRAS.462.1882K} {462, 1882}

\bibitem[\protect\citeauthoryear{{Krause}, {Chang}, {Dor{\'e}}  \&
  {Umetsu}}{{Krause} et~al.}{2013}]{2013ApJ...762L..20K}
{Krause} E.,  {Chang} T.-C.,  {Dor{\'e}} O.,   {Umetsu} K.,  2013, \mn@doi
  [ApJ] {10.1088/2041-8205/762/2/L20}, \href
  {http://adsabs.harvard.edu/abs/2013ApJ...762L..20K} {762, L20}

\bibitem[\protect\citeauthoryear{{Lares}, {Ruiz}, {Luparello}, {Ceccarelli},
  {Garcia Lambas}  \& {Paz}}{{Lares} et~al.}{2017}]{2017MNRAS.468.4822L}
{Lares} M.,  {Ruiz} A.~N.,  {Luparello} H.~E.,  {Ceccarelli} L.,  {Garcia
  Lambas} D.,   {Paz} D.~J.,  2017, \mn@doi [MNRAS] {10.1093/mnras/stx825},
  \href {http://adsabs.harvard.edu/abs/2017MNRAS.468.4822L} {468, 4822}

\bibitem[\protect\citeauthoryear{{Lewis}, {Challinor}  \& {Lasenby}}{{Lewis}
  et~al.}{2000}]{2000ApJ...538..473L}
{Lewis} A.,  {Challinor} A.,   {Lasenby} A.,  2000, \mn@doi [ApJ]
  {10.1086/309179}, \href {http://adsabs.harvard.edu/abs/2000ApJ...538..473L}
  {538, 473}

\bibitem[\protect\citeauthoryear{{Mackenzie}, {Shanks}, {Bremer}, {Cai},
  {Gunawardhana}, {Kov{\'a}cs}, {Norberg}  \& {Szapudi}}{{Mackenzie}
  et~al.}{2017}]{2017arXiv170403814M}
{Mackenzie} R.,  {Shanks} T.,  {Bremer} M.~N.,  {Cai} Y.-C.,  {Gunawardhana}
  M.~L.~P.,  {Kov{\'a}cs} A.,  {Norberg} P.,   {Szapudi} I.,  2017, preprint
  (arxiv:1704.03814), \href {http://adsabs.harvard.edu/abs/2017arXiv170403814M}
  {}

\bibitem[\protect\citeauthoryear{{Mandelbaum} et~al.,}{{Mandelbaum}
  et~al.}{2017}]{2017arXiv170506745M}
{Mandelbaum} R.,  et~al., 2017, preprint (arXiv:1705.06745), \href
  {http://adsabs.harvard.edu/abs/2017arXiv170506745M} {}

\bibitem[\protect\citeauthoryear{{Melchior}, {Sutter}, {Sheldon}, {Krause}  \&
  {Wandelt}}{{Melchior} et~al.}{2014}]{2014MNRAS.440.2922M}
{Melchior} P.,  {Sutter} P.~M.,  {Sheldon} E.~S.,  {Krause} E.,   {Wandelt}
  B.~D.,  2014, \mn@doi [MNRAS] {10.1093/mnras/stu456}, \href
  {http://adsabs.harvard.edu/abs/2014MNRAS.440.2922M} {440, 2922}

\bibitem[\protect\citeauthoryear{{Miyazaki} et~al.,}{{Miyazaki}
  et~al.}{2006}]{2006SPIE.6269E...9M}
{Miyazaki} S.,  et~al., 2006, in Society of Photo-Optical Instrumentation
  Engineers (SPIE) Conference Series. , \mn@doi{10.1117/12.672739}

\bibitem[\protect\citeauthoryear{{Nadathur}, {Lavinto}, {Hotchkiss}  \&
  {R{\"a}s{\"a}nen}}{{Nadathur} et~al.}{2014}]{2014PhRvD..90j3510N}
{Nadathur} S.,  {Lavinto} M.,  {Hotchkiss} S.,   {R{\"a}s{\"a}nen} S.,  2014,
  \mn@doi [Phys. Rev. D] {10.1103/PhysRevD.90.103510}, \href
  {http://adsabs.harvard.edu/abs/2014PhRvD..90j3510N} {90, 103510}

\bibitem[\protect\citeauthoryear{{Naidoo}, {Benoit-L{\'e}vy}  \&
  {Lahav}}{{Naidoo} et~al.}{2016}]{2016MNRAS.459L..71N}
{Naidoo} K.,  {Benoit-L{\'e}vy} A.,   {Lahav} O.,  2016, \mn@doi [MNRAS]
  {10.1093/mnrasl/slw043}, \href
  {http://adsabs.harvard.edu/abs/2016MNRAS.459L..71N} {459, L71}

\bibitem[\protect\citeauthoryear{{Planck Collaboration} et~al.,}{{Planck
  Collaboration} et~al.}{2014}]{2014A&A...571A..23P}
{Planck Collaboration} et~al., 2014, \mn@doi [A\&A]
  {10.1051/0004-6361/201321534}, \href
  {http://adsabs.harvard.edu/abs/2014A%26A...571A..23P} {571, A23}

\bibitem[\protect\citeauthoryear{{Planck Collaboration} et~al.,}{{Planck
  Collaboration} et~al.}{2016}]{2016A&A...594A..16P}
{Planck Collaboration} et~al., 2016, \mn@doi [A\&A]
  {10.1051/0004-6361/201526681}, \href
  {http://adsabs.harvard.edu/abs/2016A%26A...594A..16P} {594, A16}

\bibitem[\protect\citeauthoryear{{Rees} \& {Sciama}}{{Rees} \&
  {Sciama}}{1968}]{1968Natur.217..511R}
{Rees} M.~J.,  {Sciama} D.~W.,  1968, \mn@doi [Nat] {10.1038/217511a0}, \href
  {http://adsabs.harvard.edu/abs/1968Natur.217..511R} {217, 511}

\bibitem[\protect\citeauthoryear{Sakai \& Inoue}{Sakai \&
  Inoue}{2008}]{Sakai08}
Sakai N.,  Inoue K.~T.,  2008, \mn@doi [Phys. Rev. D]
  {10.1103/PhysRevD.78.063510}, \href
  {http://adsabs.harvard.edu/abs/2008PhRvD..78f3510S} {78, 063510}

\bibitem[\protect\citeauthoryear{{S{\'a}nchez} et~al.,}{{S{\'a}nchez}
  et~al.}{2017}]{2017MNRAS.465..746S}
{S{\'a}nchez} C.,  et~al., 2017, \mn@doi [MNRAS] {10.1093/mnras/stw2745}, \href
  {http://ads.nao.ac.jp/abs/2017MNRAS.465..746S} {465, 746}

\bibitem[\protect\citeauthoryear{{Sheth} \& {van de Weygaert}}{{Sheth} \& {van
  de Weygaert}}{2004}]{2004MNRAS.350..517S}
{Sheth} R.~K.,  {van de Weygaert} R.,  2004, \mn@doi [MNRAS]
  {10.1111/j.1365-2966.2004.07661.x}, \href
  {http://adsabs.harvard.edu/abs/2004MNRAS.350..517S} {350, 517}

\bibitem[\protect\citeauthoryear{{Shirasaki}, {Hamana}  \&
  {Yoshida}}{{Shirasaki} et~al.}{2015}]{2015MNRAS.453.3043S}
{Shirasaki} M.,  {Hamana} T.,   {Yoshida} N.,  2015, \mn@doi [MNRAS]
  {10.1093/mnras/stv1854}, \href
  {http://adsabs.harvard.edu/abs/2015MNRAS.453.3043S} {453, 3043}

\bibitem[\protect\citeauthoryear{{Springel} et~al.,}{{Springel}
  et~al.}{2005}]{2005Natur.435..629S}
{Springel} V.,  et~al., 2005, \mn@doi [nat] {10.1038/nature03597}, \href
  {http://adsabs.harvard.edu/abs/2005Natur.435..629S} {435, 629}

\bibitem[\protect\citeauthoryear{{Sugai} et~al.,}{{Sugai}
  et~al.}{2012}]{2012SPIE.8446E..0YS}
{Sugai} H.,  et~al., 2012, in Ground-based and Airborne Instrumentation for
  Astronomy IV. p. 84460Y (\mn@eprint {arXiv} {1210.2719}),
  \mn@doi{10.1117/12.926954}

\bibitem[\protect\citeauthoryear{{Sugai} et~al.,}{{Sugai}
  et~al.}{2014}]{2014SPIE.9147E..0TS}
{Sugai} H.,  et~al., 2014, in Ground-based and Airborne Instrumentation for
  Astronomy V. p. 91470T (\mn@eprint {arXiv} {1408.2825}),
  \mn@doi{10.1117/12.2054294}

\bibitem[\protect\citeauthoryear{{Szapudi} et~al.,}{{Szapudi}
  et~al.}{2015}]{2015MNRAS.450..288S}
{Szapudi} I.,  et~al., 2015, \mn@doi [MNRAS] {10.1093/mnras/stv488}, \href
  {http://adsabs.harvard.edu/abs/2015MNRAS.450..288S} {450, 288}

\bibitem[\protect\citeauthoryear{{Takahashi}, {Hamana}, {Shirasaki},
  {Namikawa}, {Nishimichi}, {Osato}  \& {Shiroyama}}{{Takahashi}
  et~al.}{2017}]{2017arXiv170601472T}
{Takahashi} R.,  {Hamana} T.,  {Shirasaki} M.,  {Namikawa} T.,  {Nishimichi}
  T.,  {Osato} K.,   {Shiroyama} K.,  2017, preprint (arXiv:1706.01472), \href
  {http://adsabs.harvard.edu/abs/2017arXiv170601472T} {}

\bibitem[\protect\citeauthoryear{{Tamura} et~al.,}{{Tamura}
  et~al.}{2016}]{2016SPIE.9908E..1MT}
{Tamura} N.,  et~al., 2016, in Ground-based and Airborne Instrumentation for
  Astronomy VI. p. 99081M (\mn@eprint {arXiv} {1608.01075}),
  \mn@doi{10.1117/12.2232103}

\bibitem[\protect\citeauthoryear{{Taylor}}{{Taylor}}{2001}]{2001astro.ph.11605T}
{Taylor} A.~N.,  2001, preprint (arXiv:0111605), \href
  {http://adsabs.harvard.edu/abs/2001astro.ph.11605T} {}

\bibitem[\protect\citeauthoryear{Tomita \& Inoue}{Tomita \&
  Inoue}{2008}]{Tomita08}
Tomita K.,  Inoue K.~T.,  2008, \mn@doi [\prd] {10.1103/PhysRevD.77.103522},
  \href {http://adsabs.harvard.edu/abs/2008PhRvD..77j3522T} {77, 103522}

\bibitem[\protect\citeauthoryear{{Vielva}, {Mart{\'{\i}}nez-Gonz{\'a}lez},
  {Barreiro}, {Sanz}  \& {Cay{\'o}n}}{{Vielva}
  et~al.}{2004}]{2004ApJ...609...22V}
{Vielva} P.,  {Mart{\'{\i}}nez-Gonz{\'a}lez} E.,  {Barreiro} R.~B.,  {Sanz}
  J.~L.,   {Cay{\'o}n} L.,  2004, \mn@doi [ApJ] {10.1086/421007}, \href
  {http://adsabs.harvard.edu/abs/2004ApJ...609...22V} {609, 22}

\bibitem[\protect\citeauthoryear{{Zhang} \& {Huterer}}{{Zhang} \&
  {Huterer}}{2010}]{2010APh....33...69Z}
{Zhang} R.,  {Huterer} D.,  2010, \mn@doi [Astroparticle Physics]
  {10.1016/j.astropartphys.2009.11.005}, \href
  {http://adsabs.harvard.edu/abs/2010APh....33...69Z} {33, 69}

\bibitem[\protect\citeauthoryear{{van Waerbeke}}{{van
  Waerbeke}}{2000}]{2000MNRAS.313..524V}
{van Waerbeke} L.,  2000, MNRAS, \href
  {http://adsabs.harvard.edu/abs/2000MNRAS.313..524V} {313, 524}

\makeatother
\end{thebibliography}







\bsp	
\label{lastpage}
\end{document}